\documentclass[twocolumn,prl,amsmath,amssymb,superscriptaddress]{revtex4} 
\usepackage{graphicx,nicefrac,pdfpages}
\usepackage[draft]{todonotes}
\usepackage{hyperref}
\hypersetup{
     colorlinks   = true,
     citecolor    = blue,
     linkcolor    = blue,
     filecolor    = blue,
     urlcolor    = blue
}

\usepackage{xr}
\usepackage{amsmath}

\begin{document} 

\title{Periodic versus Intermittent Adaptive Cycles in Quasispecies Coevolution} 

\author{Alexander Seeholzer}\altaffiliation{Present address: Laboratory of Computational Neuroscience, EPF Lausanne, 1015 Lausanne, Switzerland.}
\affiliation{Arnold-Sommerfeld-Center f\"ur Theoretische Physik and Center for NanoScience Ludwig-Maximilians-Universit\"at M\"unchen, Theresienstrasse~37, 80333 M\"unchen, Germany} 

\author{Erwin Frey}\affiliation{Arnold-Sommerfeld-Center f\"ur Theoretische Physik and Center for NanoScience Ludwig-Maximilians-Universit\"at M\"unchen, Theresienstrasse~37, 80333 M\"unchen, Germany} 

\author{Benedikt Obermayer}\email{benedikt.obermayer@mdc-berlin.de}\altaffiliation{Present address: Max-Delbr{\"u}ck-Center for Molecular Medicine, 13092 Berlin, Germany.}\affiliation{Arnold-Sommerfeld-Center f\"ur Theoretische Physik and Center for NanoScience Ludwig-Maximilians-Universit\"at M\"unchen, Theresienstrasse~37, 80333 M\"unchen, Germany}

\begin{abstract}
We study an abstract model for the coevolution between mutating viruses and the adaptive immune system. In sequence space, these two populations are localized around transiently dominant strains. Delocalization or error thresholds exhibit a novel interdependence because immune response is conditional on the viral attack. An evolutionary chase is induced by stochastic fluctuations and can occur via periodic or intermittent cycles. Using simulations and stochastic analysis, we show how the transition between these two dynamic regimes depends on mutation rate, immune response, and population size.
\end{abstract} 

\maketitle  

Evolution is commonly pictured as a dynamic process on a fitness landscape in sequence space. In general, this landscape depends not only on the genotype but varies dynamically as a function of the environment and coevolving interaction partners~\cite{Wright1932}. Prominent biological examples are the coevolutionary dynamics between the adaptive immune system and virus populations such as HIV~\cite{Nowak1990,Woo2012} or influenza~\cite{Bedford2011}, or between bacteria and their phages~\cite{Levin2013}. Continuous evolutionary innovations allow the virus to transiently escape immune suppression, triggering subsequent adaptations of the immune system. These dynamics can lead to coevolutionary cycles, which have been generally described in two different forms~\cite{Woolhouse2002,Bedford2011}: either as an intermittent series of quasistationary states connected by stochastic jumps, or as periodic and largely deterministic oscillations. From a modeling perspective, this highly complex process is determined by three main features~\cite{Perelson2002}. First, mutation rates are high and populations are large, which implies large genetic heterogeneity within the populations~\cite{Domingo1985}. This has often been pictured in terms of broad quasispecies distributions around peaks in the fitness landscape~\cite{Wang2006, Eigen1977}. At the same time, continuous adaption and coevolutionary arms races are driven by strong ecological interactions~\cite{Woolhouse2002,Tellier2013}. These modulate effective fitness landscapes~\cite{Gavrilets1998, Nowak1990,Arnoldt2012} and lead to nontrivial nonlinear population dynamics. Finally, stochastic effects in finite populations become especially pronounced whenever the first two issues are relevant at the same time~\cite{Tsimring1996, Rouzine2001, Tellier2013, Gokhale2014}.

\begin{figure}
\includegraphics[width=8.6cm]{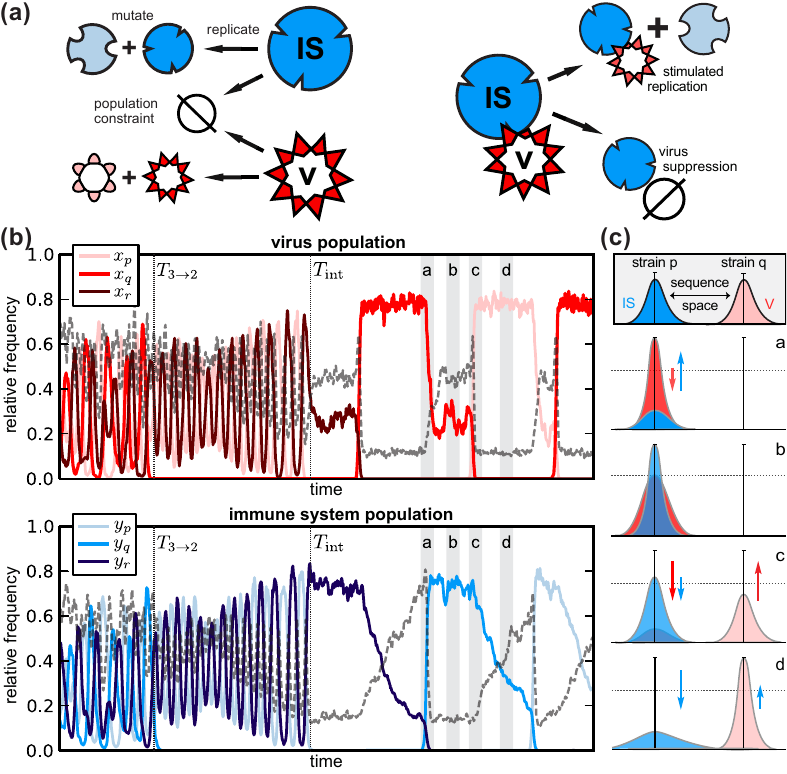}
\caption{(a) Schematic model for the coevolutionary dynamics of virus (V) and immune system (IS). The two populations are subject to mutation and selection (left), but also to ecological interactions (right).  (b) Exemplary trajectories of the relative frequencies of virulent strains $x_i$ (top) and corresponding antibodies $y_i$ (bottom). Regular oscillatory dynamics involving three strains turn into simpler two-strain oscillations at $T_{3\to2}$ and finally transition into intermittency at $T_\text{int}$. Genetic variability within the populations is calculated from the average pairwise Hamming distance and indicated in gray dashed lines. (c) Sketch of the dynamics of the full population distributions in sequence space as described in the text.}
\label{fig1}
\end{figure}

Here, we offer a synthetic perspective on these processes.
In our model [see Fig. \hyperref[fig1]{\ref*{fig1}(a)}], we consider a population of $N$ viruses represented by their genotypes (binary sequences of length $L$ and frequency $x_i$) and replication rates $r_i=1$. A small number $n$ of these genotypes corresponding to particularly virulent strains have a fitness advantage $\alpha$ over the unit baseline, giving $r_i=1+\alpha$ for $i=p,q,\ldots$. Offspring sequences undergo mutations with per-base rate $\mu_x$. In the absence of immune suppression, and in the stationary state, the viral population localizes as so-called quasispecies around any of the fittest genotypes, provided the mutation rate is smaller than Eigen's error threshold $\mu_\text{c}\approx \ln(\alpha+1)/L$~\cite{Eigen1977}. 
This simple picture is considerably complicated by the host's adaptive immune system, which produces antibodies that recognize and neutralize viruses with matching epitopes~\cite{Alberts2002}. 
Antibody production is specifically increased and variability in the binding affinity is introduced when viruses with matching genotype are encountered~\cite{Rajewsky1996}, in a process that can be modeled in terms of mutation and selection. Similar concepts can be used for bacterial immune systems, where spacer sequences in the host genome complementary to genetic elements of a phage take antibodylike functions~\cite{Levin2013}. Hence, for the immune system we introduce a second population of $N$ binary sequences with frequencies $y_i$, mutation rate $\mu_y$, unit replication rate for unstimulated production, and stimulated antibody production in the presence of perfectly matching viruses~\cite{Kamp2002, Wang2006}. Ecological interactions then introduce frequency-dependent fitness terms $\propto x_i y_i$ for such matched virus and antibody pairs. Including these terms leads to a reduction of the viral load and stimulation of antibody production [Fig. \hyperref[fig1]{\ref*{fig1}(a)}, right]. 
In the deterministic limit ($N\to\infty$), our model is described by~\cite{supplement}
\begin{equation}\label{eq:1}
\begin{split}
\dot x_i &= \textstyle{\sum_j} m^x_{ij} r_j x_j - \alpha x_i y_i - x_i\phi_x,\\
\dot y_i &= \textstyle{\sum_j} m^y_{ij} (1+\gamma x_j) y_j -y_i\phi_y,
\end{split}
\end{equation}
where $i$ and $j$ run over all $2^L$ sequences. The fitness advantage $\alpha$ of virulent strains can  be suppressed to background levels by a perfectly adapted immune system, which undergoes stimulated production at rate $\gamma$ when encountering matching viral epitopes. Further, $m^{x}_{ij}=\mu_{x}^{d_{ij}}(1-\mu_{x})^{L-d_{ij}}$ is the probability of having $d_{ij}$ simultaneous mutations, where $d_{ij}$ is the Hamming distance between $x_i$ and $x_j$. The dilution terms $\phi_{x/y}$ are obtained from the conditions $\sum_i \dot x_i=0$ and $\sum_i \dot y_i=0$, respectively, and keep the sizes of the two populations fluctuating around constant values. This constraint applies to the stationary phase of the adaptive race, while we ignore some of the effects of a changing viral load~\cite{Nowak1990,Woo2012,Wang2006,Bull2007} and also neglect immune system memory~\cite{Bianconi2010} and unspecific recognition~\cite{Alberts2002}.

To facilitate a systematic study of the effects of demographic noise by means of simulations and theoretical analysis, our starting point is the underlying stochastic master equation \cite{supplement} which has rarely been used in this context. Its deterministic limit leads to Eq.~\eqref{eq:1} and connects to established quasispecies theory~\cite{Eigen1977,Kamp2002}. Exemplary simulation results obtained with the Gillespie algorithm~\cite{Gillespie1992} are shown in Fig. \hyperref[fig1]{\ref*{fig1}(b)}, with parameters in the coexistence regime discussed below. We readily identify characteristics of the intermittent coevolutionary dynamics. First, a particularly virulent strain with its associated quasispecies ``cloud'' of mutants triggers a specific immune response (a), leading to a corresponding localization in the antibody sequence space (b). This gives alternative viral strains that are not under immune attack a fitness advantage, and after a brief ``search'' period during which the viral population becomes delocalized, this new fitness peak is colonized in a ``growth'' phase (c), awaiting the adaptive immune response (d). The delocalization and relocalization dynamics of each population in sequence space are clearly visible as transient increases in their respective mean pairwise Hamming distances [Fig. \hyperref[fig1]{\ref*{fig1}(b)}]. Intriguingly, this sequence of events can occur both in the form of regular oscillations as well as by means of stochastically intermittent cycles~\cite{Woolhouse2002}. The former occurs when the large genetic diversity within the population extends across the valleys between different fitness peaks and signifies periodic shifts in the extent to which these peaks are populated~\cite{Gavrilets1998}. The latter case indicates that adaptation proceeds stochastically via the random discovery of previously unpopulated fitness peaks by relatively tightly localized populations.

\begin{figure}
\includegraphics[width=8.6cm]{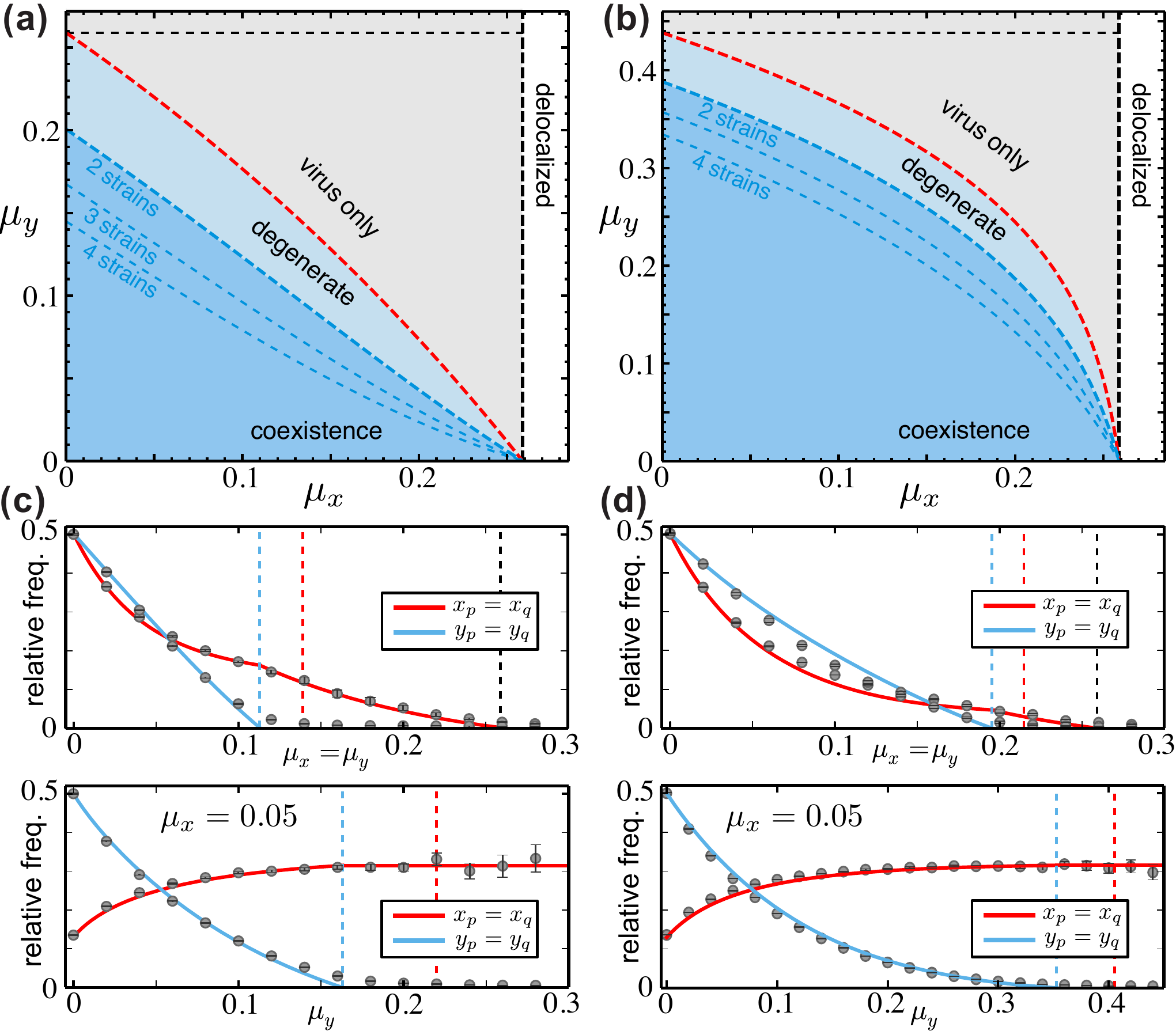}
\caption{(a) Regimes of coevolution. High mutation rates $\mu_x$ of the virus lead to population delocalization, while for lower mutation rates a regime of coexistence emerges. Intermediate values lead to a degenerate localization regime for the virus (see text). (c) Steady-state values for relative frequencies $x_p=x_q$ and $y_p=y_q$ as a function of $\mu_y$ with $\mu_x=\mu_y$ (above) or $\mu_x=0.05$ (below). Solid lines are solutions of Eq.~\eqref{eq:2} and dots are simulation results. Panels (a) and (c) are  for $\gamma=\alpha$, while (b) and (d) show analogous result for $\gamma=10\alpha$, where $L=8$, $n=2$, and $\alpha=10$.}
\label{fig2}
\end{figure}

\paragraph{Steady-state regimes: coexistence for mutation rates below interdependent error thresholds.} We use a reduced deterministic version of the model to determine stationary states and the associated error thresholds. We restrict the analysis to the populations of the $n$ virulent strains $x_{p,q,\ldots}$ and their respective antibodies $y_{p,q,\ldots}$, and lump all mutant sequences together in the so-called error tail~\cite{Schuster1988}. The high-dimensional system \eqref{eq:1} is then reduced to~\cite{supplement}
\begin{equation}\label{eq:2}
\begin{split}
\dot x_p &= \left[Q_{x}\left(1+\alpha\right)-\alpha y_{p}-\bar{\phi}_{x}\right]x_{p},\\
\dot y_p &= \left[Q_{y}\left(1+\gamma x_{p}\right)-\bar{\phi}_{y}\right]y_{p},
\end{split}
\end{equation}
where $p$ runs over the $n$ strains which are coupled by the corresponding dilution terms $\bar\phi_{x/y}$. $Q_{x/y}=(1-\mu_{x/y})^L$ are the quality factors. A straightforward stability analysis of fixed points in this system with respect to $\mu_{x/y}$ as bifurcation parameters yields the phase diagrams of Fig. \hyperref[fig2]{\ref*{fig2}}.

As expected, we recover the classical result that the viral population localizes around a fitness peak only if
$Q_{x}>Q_{c}\equiv\left(\alpha+1\right)^{-1}$, with increasing genetic variability (i.e., the width of the population distribution)
for larger mutation rate $\mu_x$. However, antibodies are localized only (1) if their mutation rate $\mu_y$ is small enough, (2) if their production rate $\gamma$ is high enough and (3) if the virus attack is specific enough (i.e., tightly localized). These interdependent requirements are an inevitable consequence of ecological interactions, and they translate into the condition $Q_y = \left\{\left(\gamma/\alpha n \right) \left[ Q_x(\alpha +1)-1\right]+1 \right\}^{-1}$ as the analytical limit for the coexistence regime (blue dashed lines in Fig.~\hyperref[fig2]{\ref*{fig2}}). Only in this regime do we find the intriguing oscillatory dynamics shown in Fig. \hyperref[fig1]{\ref*{fig1}} that will be discussed below.
Finally, in a somewhat model-specific ``degenerate'' regime bounded by $Q_y = \left\{\left(\gamma/\alpha\right) \left[ Q_x(\alpha +1)-1\right]+1 \right\}^{-1}$, the virus population can stably localize about several fitness peaks simultaneously such that none of these quasispecies is sufficiently tight to trigger a specific response of the immune system, which thus remains delocalized. The fixed points of the approximate system~\eqref{eq:2} coincide closely with the mean steady-state concentrations obtained by stochastic simulation of the full system~\eqref{eq:1} [see Figs. \hyperref[fig2]{\ref*{fig2}(c)} and \hyperref[fig2]{\ref*{fig2}(d)}]. Interestingly, for $\alpha=\gamma$ and symmetric mutation rates $\mu_{x,y} \equiv \mu$, the critical condition of coexistence can be approximated by $\mu \approx \left(1/2L\right)\ln\left(\alpha/2\right)$ for large $\alpha$ and $L$, which generalizes a comparable result for mutualistic frequency-dependent fitness~\cite{Obermayer2009} to the case of antagonistic interactions. This correspondence also suggests that the error thresholds derived here should be largely unchanged if recognition between the two population tolerates some mismatches~\cite{Obermayer2010}.

\paragraph{Noise-driven oscillations in the coexistence regime.} Performing a linear stability analysis in the coexistence regime reveals that the oscillations seen in the simulations are caused by $n-1$ pairs of purely imaginary eigenvalues. Numerical solutions of the deterministic  Eqs.~\eqref{eq:2} show complex but regular oscillations involving all $n$ strains with slow amplitude variations controlled by higher-order nonlinearities (see Fig.~S1 in the Supplemental Material~\cite{supplement}). Results from stochastic simulations, however, suggest that such complex patterns quickly give rise to simpler oscillations involving only two strains, which at a later time transition into intermittency (cf. Fig.~\hyperref[fig1]{\ref*{fig1}}). Investigating the case $n > 2$ by simulations below, we restrict further analysis to $n=2$. Also, here we only display more compact analytical results for the case $\gamma=\alpha$ (see the Supplemental Material~\cite{supplement} for general results). Our analysis exploits that in the coexistence regime mutation rates $\mu_{x/y} \lesssim \ln\alpha/L$ are small compared to the error thresholds and can be used as expansion parameters. To obtain the nonlinearities that control oscillation amplitudes, we expand Eq.~\eqref{eq:2} to first order and transform to polar normal form on the two-dimensional stable manifold~\cite{supplement}:
\begin{subequations}\label{eq:3}
\begin{align}
\dot u &= -\frac{4}{5}L\left[\mu_{x}\left(\alpha+1\right)+\mu_{y}\right] u^2,\label{eq:3a}\\
\dot \varphi &= \frac{\alpha}{2} - \frac{L}{2}(\alpha + 1) (\mu_x + \mu_y)+\mathcal{O}(u),\label{eq:3b}
\end{align}
\end{subequations}
where $u$ is a squared radial coordinate indicating deviations from the coexistence fixed point and $\varphi$ measures the phase of the oscillations.  Equation~\eqref{eq:3a} exhibits a weak geometric decay of the oscillation amplitude $\mathcal{O}(u^2 \mu_x)\ll1$, which makes the fixed point only marginally stable and thus vulnerable to stochastic fluctuations~\cite{Reichenbach2006,Cremer2009,Bladon2010}. Notably,  the oscillation frequency of Eq.~\eqref{eq:3b} depends mainly on the fitness advantage $\alpha$, and is only weakly slowed down by mutations. In this deterministic regime, the quasispecies distribution in sequence space is broad enough that the time required to shift to a new fitness peak is dominated by the \emph{growth} of the subpopulation already on the new peak (with a rate $\alpha$) rather than the \emph{search} for this new peak in the first place (via mutations). We note that this effect is even stronger if the two fitness peaks are close in sequence space, i.e., if direct mutations between them are not ignored as in Eq.~\eqref{eq:2}. In contrast, when the coexistence regime displays intermittent dynamics, because the relevant sequence space is not already inhabited by the virus population, the dynamics are inherently stochastic and mutation rates can be \emph{too small} for the virus to explore enough sequence space to escape immune suppression in time. This would correspond to an adaptation threshold as found in a previous study~\cite{Kamp2002}. However, as shown more formally below, this situation is incompatible with the presence of deterministic dynamics, which is an assumption of standard quasispecies theory. Instead, population genetics models should be used~\cite{Tellier2013,Arnoldt2012}. 

\paragraph{Noise determines if dynamics are periodic or intermittent.}
We can characterize how stochastic noise controls the transition between periodic and intermittent adaptive dynamics by means of stochastic averaging. This technique enables a systematic derivation of effective one-dimensional Fokker-Planck equations in relevant subspaces of more complex high-dimensional nonlinear dynamics such as those arising in evolutionary game theory~\cite{Dobrinevski2012}. It is based on the time scale separation between slow radial and fast azimuthal dynamics in Eq.~\eqref{eq:3}: $\varphi$ evolves on fast time scales ($\dot\varphi\propto \alpha$), while $u$ changes much more slowly ($\dot u\propto \mu L u^2$). Using this observation, we can derive effective coefficients governing the evolution of the probability distribution $P(u,t)$ of the radial variable by averaging the angular dynamics over one oscillation period \cite{supplement}. To leading order, we get 
\begin{equation}\label{eq:4}
  \partial_{t}P=  -\partial_{u}\left[\left(-a_{1}u^{2}+\frac{a_{2}}{N}\right)P \right]  +\frac{1}{N}\partial_{u}^{2}\left(a_{2}u\,P\right),
\end{equation}
with $a_1=\tfrac{4}{5}L[\mu_{x}\left(\alpha+1\right)+\mu_{y}]$ and 
$a_2=\frac{1}{16}\big\{4+3 \alpha - \mu_x L\left[\left(4/\alpha\right)+11+7 \alpha \right]- \mu_y L\left[\left(4/\alpha\right)+7+3 \alpha\right]\big\}$.
Note that in the deterministic limit $N\to\infty$ we recover Eq.~\eqref{eq:3a}. For a finite population, we now find the deterministic decay $\propto a_1 u^2$ towards the coexistence fixed point in competition with a stochastic outward drift $\propto \left( a_2 / N \right)$, which destabilizes the fixed point and leads to a finite oscillation amplitude 
$\langle u\rangle= \sqrt{\left(2/\pi\right)\left(a_2/a_1 N\right)}$
.  As mutation rates get small, expected oscillation amplitudes grow as $[(\alpha+1) \mu_x + 4 \mu_y]^{-1/2}$, eventually hitting the borders of the concentration simplex. This indicates the transition from regular oscillations to intermittent behavior: during large-amplitude oscillations the fittest virus genotypes are temporarily lost from the population and are only much later recovered through spontaneous mutations. 

\begin{figure}
\includegraphics[width=8.6cm]{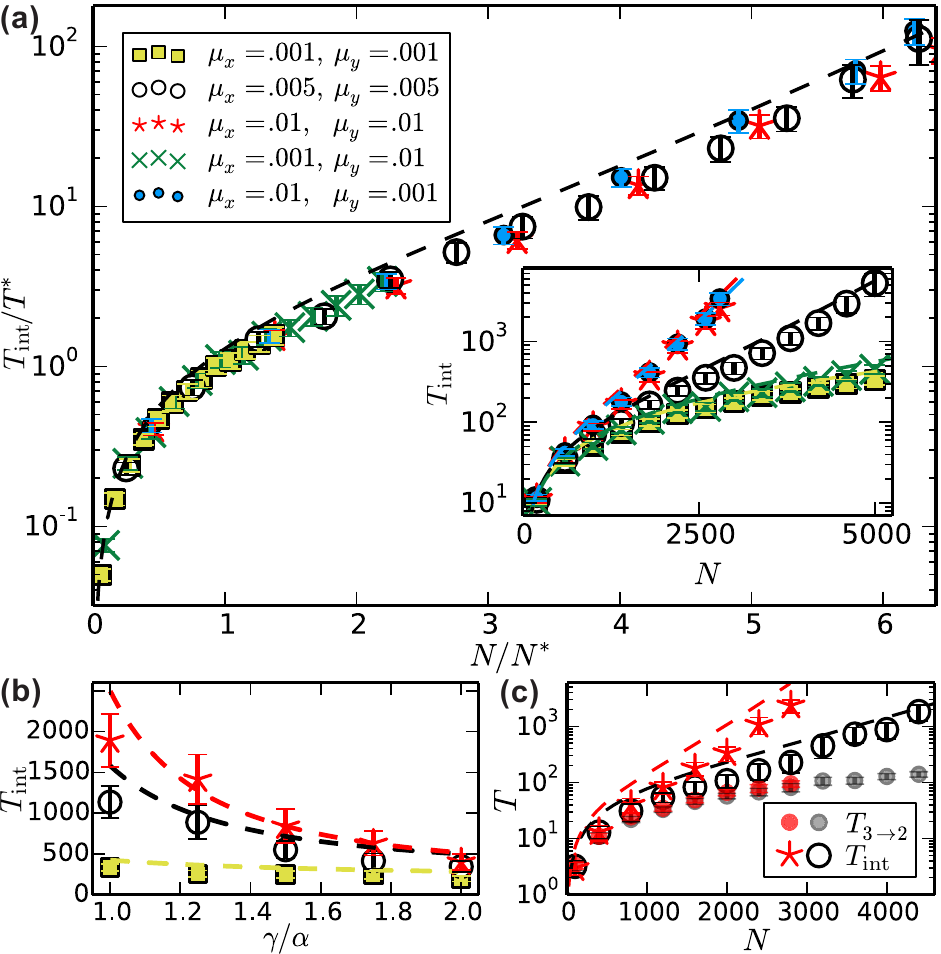}
\caption{Mean time until transition from regular oscillations to intermittency. Dashed lines are from the analytical result~\eqref{eq:5}. (a) Rescaled simulation data for $L=8$, $\alpha=\gamma=10$ and different choices of $\mu_{x/y}$ collapse onto a universal curve (unscaled data shown in the inset). (b) Transition times decrease for increasing $\gamma$ [parameters otherwise as in (a)]. (c) For $n=3$ virulent strains, the transition time $T_{3\to 2}$ until one strain is lost is much shorter than $T_\text{int}$, especially for large populations.}
\label{fig3}
\end{figure}

A more detailed understanding of this transition is obtained by estimating the lifetime of the regular oscillations. To this end, we use the bounds on the radial variable $u_\text{max} = \tfrac{1}{8}-\mathcal{O}(\mu L)$, where the populations are fully localized about only one peak. The chances of observing a transition to intermittent behavior are estimated from the mean first passage time (MFPT) $T_\text{int}$ from $u=0$ to $u=u_\text{max}$ under Eq.~\eqref{eq:4}. Using standard methods~\cite{Kampen1992}, we find the result \cite{supplement}
\begin{equation}\label{eq:5}
T_\text{int} = N \frac{u_\text{max}}{a_2} \tilde{F}\left(\frac{N a_1 u^2_\text{max}}{2 a_2}\right),
\end{equation}
where ${\tilde F}(x)$ is the generalized hypergeometric function ${_2 F_2} (\nicefrac{1}{2},1;\nicefrac{3}{2},\nicefrac{3}{2}; x)$. Equation~\eqref{eq:5} can be brought into scaling form by defining 
$N^*=\left(2 a_2/a_1 u^2_\text{max} \right)$
 and 
 $T^*=N^* \left(u_\text{max} / a_2 \right)$. To compare this result to simulations, we plot the rescaled MFPT 
 $\left(T_\text{int}/T^* \right) = \left(N/N^*\right) {\tilde F} (N/N^*)$
  (see Fig.~\hyperref[fig3]{\ref*{fig3}}). The nearly perfect data collapse for different parameter choices validates our analytical approach. 

While $N^*$ measures the population size at the crossover from periodic to intermittent dynamics, $T^*$ denotes the corresponding typical duration of the transition. For large populations ($N > N^*$), we find $T_\text{int}\sim N^{-1/2} e^{N/N^*}$; this almost exponential growth of the MFPT indicates that the dynamics are effectively deterministic and intermittent behavior extremely unlikely. For $N < N^*$, we find $T_\text{int}\sim N$ and the dynamics thus easily transition into intermittency. This distinction based on the scaling of $T_\text{int}$ with $N$ has recently been suggested in the context of game theory~\cite{Cremer2009}. In our case, however, finite mutation rates prevent permanent extinction of subpopulations and stabilize regular oscillatory behavior even in small populations, because the deterministic decay in Eq.~\eqref{eq:3a} is strengthened and the critical population size $N^* \propto (\mu_x L)^{-1}$ is reduced. Thus, even for small populations, mutations can act as a driving force for the stabilization of regular oscillations, which \emph{a posteriori} justifies assumptions underlying quasispecies theory and generalizes previous observations~\cite{Gokhale2014}. In contrast, from results for general $\gamma$~(see Fig.~S2 in the Supplemental Material~\cite{supplement}), we find that a strong immune response (i.e., $\gamma > \alpha$) promotes early transitions into intermittency [cf. Fig.~\hyperref[fig3]{\ref*{fig3}(b)}], since both $N^*$ and $T^*$ increase with $\gamma$. However, these parameters are insensitive against the precise value of $\mu_y$ [cf. Fig.~\hyperref[fig3]{\ref*{fig3}(a)}]; this suggests that effective immune suppression is achieved via a strong stimulated response rather than high adaptive flexibility. Indeed, extreme antibody secretion rates have been reported in the literature~\cite{Hibi1986}.  Finally, we support our choice of limiting the analysis to $n=2$ strains by simulating a system with $n=3$ strains, measuring the time $T_{3\to 2}$ until one strain is lost as well as the subsequent $T_\text{int}$ until the remaining two strains transition to intermittency. As shown in Fig.~\hyperref[fig3]{\ref*{fig3}(c)}, the state with all three strains present is short lived compared to the two-strain oscillations, especially in the relevant deterministic regime of larger population size. Hence, apart from numerical prefactors the general trend captured in Eq.~\eqref{eq:5} also describes systems with larger $n$. 

\paragraph{Conclusions.} We have analyzed a model for the coevolutionary dynamics of virus and immune system, combining simulations with nonlinear deterministic and stochastic analysis. Starting from the established quasispecies treatment of this problem, we explicitly introduced interactions between the populations. These lead to interdependent error thresholds, because a focused immune defense against a specific viral strain is impaired for large genetic variability in the virus population. Further, we performed a rigorous analysis of stochastic effects in the coexistence regime: regular yet noise-induced oscillatory behavior for large populations, large mutation rates, and weak immune response turn into stochastic intermittent cycles for smaller populations, smaller mutation rates and strong immune response. Our simulations indicate that the reverse transition from intermittency towards regular oscillations is a rare event occurring on time scales well beyond $T_\text{int}$. It cannot easily be analyzed within our reduced two-dimensional model as it will depend on the entire population structure. Finally, we note that our abstract model based on quasispecies theory focuses on the dynamics of genetic variability within populations of constant size. This assumption is of course violated for some biological scenarios, where immune response modulates the viral load~\cite{Nowak1990,Wang2006, Bull2007} and may well lead to extinction of the virus~\cite{Tejero2010,Woo2012}. We expect that more detailed models including these and other effects relevant in biological situations~\cite{Bianconi2010,Alberts2002} will also be amenable to theoretical analysis based on the stochastic averaging techniques used here.

\begin{acknowledgements}
We acknowledge helpful comments by anonymous reviewers and financial support by the Deutsche Forschungsgemeinschaft in the framework of the SFB/TR12 -- Symmetries and Universality in Mesoscopic Systems.
\end{acknowledgements}

\pagebreak


\section*{Supplementary material}

# Periodic versus Intermittent Adaptive Cycles in Quasispecies Coevolution: Supplemental Material

Alexander Seeholzer,[1,][*] Erwin Frey,[1] and Benedikt Obermayer[1,][†]

[1]*Arnold-Sommerfeld-Center für Theoretische Physik and Center for NanoScience,
Ludwig-Maximilians-Universität München, Theresienstr. 37, 80333 München, Germany*

**Contents**

## I. STOCHASTIC MODEL

### A. Reaction Network

Let $N$ be the desired population size for both virus and immune system (IS) populations. We define

$$m_{ij}^{x/y} = \mu_{x/y}^{d_{ij}} (1-\mu_{x/y})^{L-d_{ij}},$$

where $d_{ij}$ is the bitwise Hamming distance between sequences $\sigma_i^{x/y}$ and $\sigma_j^{x/y}$ (number of bits differing between sequences). Note that $m_{ii}^{x/y} = Q_{x/y}$, the so called *quality factor* as also defined in the main text.

Possible reactions for our network of virus particles $\sigma_i^x$ ($i \in 1,...,2^L$) with absolute population numbers $\mathbf{n}^x = (n_1^x,...,n_{2^L}^x)$ and IS particles $\sigma_i^y$ ($i \in 1,...,2^L$) with absolute population numbers $\mathbf{n}^y = (n_1^y,...,n_{2^L}^y)$ are defined as follows:

---

[*]Present address: Laboratory of Computational Neuroscience, EPF Lausanne, 1015 Lausanne, Switzerland.
[†]Electronic address: benedikt.obermayer@mdc-berlin.de; Present address: Max-Delbrück-Center for Molecular Medicine, 13092 Berlin, Germany.

1. *Virus error prone reproduction*: $\sigma_j^x \to \sigma_j^x + \sigma_i^x$ at rate (per unit time)

$$R_{ij}^x = m_{ij}^x r_j.$$

   The virus fitness landscape is defined by $r_j = 1 + \alpha$ if $\sigma_j^x$ is a virulent strain, otherwise $r_j = 1$

2. *IS error prone (stimulated) reproduction*: $\sigma_j^y \to \sigma_j^y + \sigma_i^y$ at rate (per unit time)

$$R_{ij}^y(\mathbf{n}^x) = m_{ij}^y (1 + \frac{\gamma}{N} n_j^x).$$

   Normalization by $N$ ensures the proper scaling with system size and thus the proper deterministic limit Eq. (1).

3. *Virus suppression by IS*: $\sigma_i^x \to \emptyset$ at rate (per unit time)

$$D_i(\mathbf{n}^y) = \frac{\alpha}{N} n_i^y.$$

   Normalization by $N$ as explained above.

4. *Dilution fluxes*: $\sigma_i^x \to \emptyset$ at rates (per unit time) proportional to the mean excess productions

$$\bar{R}^x(\mathbf{n}^x, \mathbf{n}^y) = \frac{1}{N} \sum_j (r_j - \frac{\alpha}{N} n_j^y) n_j^x \tag{S1}$$

$$\bar{R}^y(\mathbf{n}^x, \mathbf{n}^y) = \frac{1}{N} \sum_j (1 + \frac{\gamma}{N} n_j^y) n_j^x \tag{S2}$$

   This type of dilution reproduces the dilution flux proposed by Eigen [1] in the deterministic limit and has been shown [2] to keep population sizes fluctuating around the desired magnitude $N$. As long as $\alpha \ll N$ we can assume that $\bar{R}^x \geq 0$, especially for small mutation probabilities, since then $n_j^y \gg 1$ only if $\sigma_j^x$ is a virulent strain, for which $r_j = \alpha + 1$. For all parameter ranges and reactions used to generate simulation results for this publication (also including higher mutation probabilities) the case $\bar{R}^x < 0$ did not occur.

These reactions were implemented in the framework of Gillespie [3] to generate all realizations of the stochastic dynamics in this publication. All simulations used $\alpha = 10$ and $L = 8$.

### B. Master equation & Fokker-Planck equation

The master equation of the reaction network as defined in the last section can be straightforwardly stated as below. $\mathbf{e}_i$ is the i-th unit vector and indices are assumed to run from 1 to $2^L$ if not stated otherwise.

$$\begin{aligned}
\partial_t P(\mathbf{n}^x, \mathbf{n}^y, t) = \sum_i &\left\{ \left[ \sum_{j \neq i} R_{ij}^x n_j^x + R_{ii}^x (n_i^x - 1) \right] P(\mathbf{n}^x - \mathbf{e}_i, \mathbf{n}^y, t) \right. \\
&+ \left[ \sum_{j \neq i} R_{ij}^y(\mathbf{n}^x) n_j^y + R_{ii}^y(\mathbf{n}^x)(n_i^y - 1) \right] P(\mathbf{n}^x, \mathbf{n}^y - \mathbf{e}_i, t) \\
&+ \left[ D_i(\mathbf{n}^y) + \bar{R}^x(\mathbf{n}^x + \mathbf{e}_i, \mathbf{n}^y) \right] (n_i^x + 1) P(\mathbf{n}^x + \mathbf{e}_i, \mathbf{n}^y, t) \\
&+ \bar{R}^y(\mathbf{n}^x, \mathbf{n}^y + \mathbf{e}_i)(n_i^y + 1) P(\mathbf{n}^x, \mathbf{n}^y + \mathbf{e}_i, t) \\
&- \left[ \sum_j R_{ij}^x n_j^x + D_i(\mathbf{n}^y) n_i^x + \bar{R}^x(\mathbf{n}^x, \mathbf{n}^y) n_i^x \right] P(\mathbf{n}^x, \mathbf{n}^y, t) \\
&\left. - \left[ \sum_j R_{ij}^y(\mathbf{n}^x) n_j^y + \bar{R}^y(\mathbf{n}^x, \mathbf{n}^y) n_i^y \right] P(\mathbf{n}^x, \mathbf{n}^y, t) \right\}.
\end{aligned} \tag{S3}$$

We now derive a Fokker-Planck equation from the master equation (S3) by a Kramers-Moyal expansion [4]. We change variables from absolute numbers $\mathbf{n}^x, \mathbf{n}^y$ to concentrations $\mathbf{x} = (x_1, ..., x_{2^L}) \equiv \frac{1}{N} \mathbf{n}^{\mathbf{x}}$ and $\mathbf{y} = (y_1, ..., y_{2^L}) \equiv \frac{1}{N} \mathbf{n}^{\mathbf{y}}$. Note

that due to the scaling of the reaction rates as chosen above, the frequency dependent reaction rates above transform as

$$R_{ij}^y(\mathbf{n}^x) = m_{ij}^y(1 + \gamma x_j) \equiv R_{ij}^{\prime y}(\mathbf{x}),$$
$$D_i(\mathbf{n}^y) = \alpha y_i \equiv D_i'(\mathbf{y}),$$
$$\bar{R}^x(\mathbf{n}^x, \mathbf{n}^y) = \sum_j (r_j - \alpha y_j) x_j \equiv \phi_x(\mathbf{x}, \mathbf{y}),$$
$$\bar{R}^y(\mathbf{n}^x, \mathbf{n}^y) = \sum_j (1 + \gamma x_j) y_j \equiv \phi_y(\mathbf{x}, \mathbf{y}).$$

Denoting $\Delta = \frac{1}{N}$ and $\mathbf{\Delta}_i = \Delta \mathbf{e}_i$, the master equation (S3) now becomes:

$$\begin{aligned}\partial_t P(\mathbf{x}, \mathbf{y}, t) =& N \sum_i \Bigg\{ \left[ \sum_{j \neq i} R_{ij}^x x_j + R_{ii}^x (x_i - \Delta) \right] P(\mathbf{x} - \mathbf{\Delta}_i, \mathbf{y}, t) \\ &+ \left[ \sum_{j \neq i} R_{ij}^{\prime y}(\mathbf{x}) y_j + R_{ii}^{\prime y}(\mathbf{x})(y_i - \Delta) \right] P(\mathbf{x}, \mathbf{y} - \mathbf{\Delta}_i, t) \\ &+ [D_i'(\mathbf{y}) + \phi_x(\mathbf{x} + \mathbf{\Delta}_i, \mathbf{y})](x_i + \Delta) P(\mathbf{x} + \mathbf{\Delta}_i, \mathbf{y}, t) \\ &+ \phi_y(\mathbf{x}, \mathbf{y} + \mathbf{\Delta}_i)(y_i + \Delta) P(\mathbf{x}, \mathbf{y} + \mathbf{\Delta}_i, t) \\ &- \left[ \sum_j R_{ij}^x x_j + D_i'(\mathbf{y}) x_i + \phi_x(\mathbf{x}, \mathbf{y}) x_i \right] P(\mathbf{x}, \mathbf{y}, t) \\ &- \left[ \sum_j R_{ij}^{\prime y}(\mathbf{x}) y_j + \phi_y(\mathbf{x}, \mathbf{y}) y_i \right] P(\mathbf{x}, \mathbf{y}, t) \Bigg\}.\end{aligned}$$

For $N$ large enough we can treat $\mathbf{x}$ and $\mathbf{y}$ as continuous variables. The Kramers-Moyal expansion then consists of an expansion of the right-hand side up to $\Delta^2$. This then yields the Fokker-Planck equation ($\delta_{ik} = 1$ if $i = k$ and 0 otherwise)

$$\begin{aligned}\partial_t P(\mathbf{x}, \mathbf{y}, t) =& -\sum_i \{\partial_{x_i}[A_i^x P(\mathbf{x}, \mathbf{y}, t)] + \partial_{y_i}[A_i^y P(\mathbf{x}, \mathbf{y}, t)]\} \\ &+ \frac{1}{2N} \sum_{i,k} \{\partial_{x_i}\partial_{x_k}[B_{ik}^x P(\mathbf{x}, \mathbf{y}, t)] + \partial_{y_i}\partial_{y_k}[B_{ik}^y P(\mathbf{x}, \mathbf{y}, t)]\}, \quad (S4) \\ A_i^x =& \sum_j m_{ij}^x r_j x_j - \alpha x_i y_i - x_i \phi_x(\mathbf{x}, \mathbf{y}), \\ A_i^y =& \sum_j m_{ij}^y (1 + \gamma x_j) y_j - y_i \phi_y(\mathbf{x}, \mathbf{y}), \\ B_{ik}^x =& \delta_{ik} \sum_j m_{ij}^x r_j x_j + \alpha x_i y_i + x_i \phi_x(\mathbf{x}, \mathbf{y}), \\ B_{ik}^y =& \delta_{ik} \sum_j m_{ij}^y (1 + \gamma x_j) y_j + y_i \phi_y(\mathbf{x}, \mathbf{y}).\end{aligned}$$

These correspond to a set of $2 \cdot 2^L$ coupled nonlinear Itō stochastic differential equations (SDE) [4]:

$$\begin{aligned}dx_i &= A_i^x dt + \frac{1}{\sqrt{N}} C_{ii}^x dW_i^x, \quad i \in \{1, ..., 2^L\} \\ dy_i &= A_i^y dt + \frac{1}{\sqrt{N}} C_{ii}^y dW_i^y, \quad i \in \{1, ..., 2^L\},\end{aligned} \quad (S5)$$

where $C^{x/y}$ is a diagonal matrix satisfying $B_{ii}^{x/y} = \left(C_{ii}^{x/y}\right)^2$ and $W_i^{x/y}$ are independent Wiener processes with zero mean and unit variance. We thus see that in the deterministic limit $N \to \infty$ we recover the deterministic equations of the main text Eqs. (1).

### C. Mean pairwise Hamming distance

For a given distribution $\mathbf{n} = (n_1, ..., n_{2^L})$ of sequences in a population, the mean pairwise Hamming distance is defined by

$$d_{\text{pairwise}} = \left(\sum_{i=0}^{2^L} n_i\right)^{-2} \sum_{i,j=0}^{2^L} n_i n_j d_{ij},$$

where $d_{ij}$ is the bitwise Hamming distance between sequences $\sigma_i^{x/y}$ and $\sigma_j^{x/y}$ (number of bits differing between sequences). For the gray dashed lines displayed in Fig. Fig. 1(b) of the main text, the value of $d_{pw}$ is further normalized by the maximal mean pairwise Hamming distance of $\frac{L}{2}$ (for a uniform distribution of sequences).

## II. DETERMINISTIC APPROXIMATION

### A. Error-tail approximation

We will simplify the high dimensional system (S5) by applying the error-tail approximation [5, 7], under the simple virus fitness landscape of only a few virulent strains $x_p, x_q, x_r, ...$. In the case of only two virulent strains $x_p$ is the concentration of sequence $(0, ..., 0)$ and $x_q$ is the concentration of sequence $(1, ..., 1)$. We define $\mathbb{S} = p, q, r, ...$ to be the set indices of virulent strains. For a start we restrict the analysis to the equations for $x_i, i \in \mathbb{S}$ and the matching immune system (IS) sequences $y_i, i \in \mathbb{S}$ only. Let the *error-tails* of the respective populations be $x_e = \sum_{k \notin \mathbb{S}} x_k = 1 - \sum_{i \in \mathbb{S}} x_i$ (note that $\sum_{k=1}^{2^L} x_k = 1$) and $y_e$ accordingly. The restricted system (S5) then reads:

$$\begin{aligned}
dx_i &= A_i^x dt + \frac{1}{\sqrt{N}} C_{ii}^x dW_i^x, \\
dy_i &= A_i^y dt + \frac{1}{\sqrt{N}} C_{ii}^y dW_i^y,
\end{aligned} \quad (S6)$$

for indices $i \in \mathbb{S}$.

The error-tail approximation now consists of considering only mutations from the virulent strains and matching IS sequences into the error tail, explicitly neglecting back mutations. While this approximation is analytically valid only for $L \to \infty$ and $\mu_x, \mu_y \to 0$, it has been successfully applied even for relatively short lengths $L$ and larger mutation probabilities [6]. Since the Hamming distance between the considered sequences is maximal, we can also neglect mutations between them. With these considerations the coefficients read ($i \in \mathbb{S}$):

$$\begin{aligned}
A_i^x &= \left[Q_x(1+\alpha) - \alpha y_i - \bar{\phi}_x\right] x_i, \\
(C_{ii}^x)^2 &= \left[Q_x(1+\alpha) + \alpha y_i + \bar{\phi}_x\right] x_i, \\
A_i^y &= \left[Q_y(1+\gamma x_i) - \bar{\phi}_y\right] y_i, \\
(C_{ii}^y)^2 &= \left[Q_y(1+\gamma x_i) + \bar{\phi}_y\right] y_i.
\end{aligned} \quad (S7)$$

Here we have again introduced the *quality-factor* $Q_{x/y} = (1 - \mu_{x/y})^L$. The error-tail approximated dilution fluxes are given below – there, frequency dependent fitness terms in the error-tail, i.e. terms $\sim x_i y_i$ for $i \notin \mathbb{S}$, are neglected:

$$\begin{aligned}
\bar{\phi}_x &= \sum_{i \in \mathbb{S}} (1 + \alpha - \alpha y_i) x_i + x_e = 1 + \alpha \sum_{i \in \mathbb{S}} (1 - y_i) x_i, \\
\bar{\phi}_y &= \sum_{i \in \mathbb{S}} (1 + \gamma x_i) y_i + y_e = 1 + \gamma \sum_{i \in \mathbb{S}} x_i y_i.
\end{aligned}$$

In the deterministic limit $N \to \infty$ the $2 * n$ dimensional system (S6) with coefficients (S7) gives the reduced system Eqs. (2) of the main text.

### B. Coexistence fixpoints and bifurcations

In the coexistence regime introduced in the main text, the system of equations (S6) admits the following set of fixed points in the deterministic limit ($N \to \infty$). Let the number of virulent strains be $n \equiv |\mathbb{S}|$, then the coexistence

fixed point is given by ($i \in \mathbb{S}$)

$$
\begin{aligned}
x_i = x_m &\equiv \frac{Q_y\left(1 - \frac{\gamma}{n}\right) + Q_x\left(\frac{\gamma}{\alpha} + \gamma\right) - \frac{\gamma}{\alpha} - 1 + \sqrt{D}}{2\gamma(n - Q_y)}, \\
y_i = y_m &\equiv \frac{Q_y\left(1 + \frac{\gamma}{n}\right) + Q_x\left(\frac{\gamma}{\alpha} + \gamma\right) - \frac{\gamma}{\alpha} - 1 - \sqrt{D}}{2\gamma}, \\
D &= \frac{4\gamma}{n}(1 - Q_y)(n - Q_y) + \left[Q_y\left(1 - \frac{\gamma}{n}\right) + Q_x\left(\gamma + \frac{\gamma}{\alpha}\right) - 1 - \frac{\gamma}{\alpha}\right]^2.
\end{aligned} \tag{S8}
$$

If $Q_y < \left[\frac{\gamma}{\alpha n}(Q_x(\alpha + 1) - 1) + 1\right]^{-1}$ the IS coexistence solution $y_m$ becomes negative, unstable and $y_i = 0$ for $i \in \mathbb{S}$ becomes the stable fixed point for the IS. The total virus concentration $x_{\max} = \sum_{i \in \mathbb{S}} x_i$ is then restricted by the virus mutation rate via $x_{\max} = \frac{Q_x(\alpha+1)-1}{\alpha}$. For $Q_x = Q_c$ the analytical prediction of the virus concentration vanishes, $x_{\max} = 0$, which is the transition into the delocalized regime (see main text), where now also $x_i = 0$ for $i \in \mathbb{S}$.

While the solutions $x_{\max} = \sum_{i \in \mathbb{S}} x_i$ and $y_i = 0$ for $i \in \mathbb{S}$ represents a line of fixed points, not the whole line is stable. For $n = 2$ it can easily be shown (by evaluating the Jacobian of the system (S6) for $N \to \infty$) that for $\left[\frac{\gamma}{\alpha}(Q_x(\alpha + 1) - 1) + 1\right]^{-1} \leq Q_y \leq \left[\frac{\gamma}{2\alpha}(Q_x(\alpha + 1) - 1) + 1\right]^{-1}$ (i.e. in the degenerate regime of Fig. 2) the stable segment of the line is given by

$$
\begin{aligned}
x_p &= \frac{x_{\max}}{2}(1 \pm \Delta), \\
x_q &= \frac{x_{\max}}{2}(1 \mp \Delta), \\
\Delta &\leq \frac{2(1 - Q_y)}{Q_y(Q_x(\alpha + 1) - 1)}\frac{\alpha}{\gamma} - 1.
\end{aligned} \tag{S9}
$$

For $Q_y = \left[\frac{\gamma}{2\alpha}(Q_x(\alpha + 1) - 1) + 1\right]^{-1}$ (blue dashed lines in Fig. 2) it holds that $\Delta = 0$ and only the point $x_p = x_q$ is stable. As now either $Q_x$ or $Q_y$ are decreased (by increasing the mutation probabilities $\mu_x, \mu_y$) the size of the line of degenerate stable fixed points increases until at $Q_y = \left[\frac{\gamma}{\alpha}(Q_x(\alpha + 1) - 1) + 1\right]^{-1}$ (red dashed lines in Fig. 2) it holds that $\Delta = 1$ and all combinations of virus concentrations with $x_{\max} = x_p + x_q$ are (meta) stable.

In the insets of Fig. 2 the red solid lines show the stable branch of fixed points given by Eq. (S8) to the left of the dashed blue vertical line. To the right of the dashed blue vertical line we plot in solid red the center of the the degenerate line of fixed points $x_p = x_q = \frac{x_{\max}}{2}$. Dots display the average and 95% confidence interval of the temporal means of 40 concentration trajectories ($N = 2500$). Due to the symmetry of concentrations, only trajectories of $x_p$ and $y_p$ are used. Note that due to fluctuations in stochastic simulations the temporal mean of concentrations stays very close to the line $x_p = x_q$ even in the virus only regime, although the variability across runs increases.

### C. Normal form

#### 1. Eigenvalues of coexistence fixed point

To investigate the deterministic stability of the system (S6), we consider the eigenvalues of its Jacobian at the coexistence fixed point (cf. Eqs. (S8)). It can be readily verified for small $n$ that the eigenvalues are given by:

$$
\begin{aligned}
\lambda_{s_{1/2}} &= \frac{1}{2}\left[-2nx(\alpha + y(\gamma - \alpha)) + (\alpha + 1)Q_x + Q_y(\gamma x + 1) - \alpha y - 2 \pm \sqrt{D_s}\right], \\
\lambda_{c_{k,1}} &= \bar{\lambda}_{c_{k,2}} = \frac{1}{2}\left[-nx(\alpha + y(\gamma - \alpha)) + (\alpha + 1)Q_x + Q_y(\gamma x + 1) - \alpha y - 2 \pm \sqrt{D_c}\right], \\
D_s &= 2\gamma x\left[2\alpha n^2 x(y - 2)y - (\alpha + 1)Q_x(Q_y - 2ny) - Q_y(-2\alpha nx + 2ny + \alpha y) + Q_y^2\right] \\
&\quad + \left[\alpha(2nx(y - 1) + Q_x - y) + Q_x - Q_y\right]^2 + \gamma^2 x^2(Q_y - 2ny)^2, \\
D_c &= -2\gamma x\left[(\alpha + 1)Q_x(Q_y - ny) + \alpha Q_y(nx(y - 1) + y) + nQ_y y + \alpha ny(-nxy + nx + y) - Q_y^2\right] \\
&\quad + \left[\alpha(nx(y - 1) + Q_x - y) + Q_x - Q_y\right]^2 + \gamma^2 x^2(Q_y - ny)^2.
\end{aligned} \tag{S10}
$$

where $k \in \{1, ..., n-1\}$. For parameters in the coexistence phase, $\lambda_{s_{1/2}}$ are 2 eigenvalues with negative real parts and $\lambda_{c_{k,1}}$ and $\lambda_{c_{k,2}}$ are $2(n-1)$ pairs of conjugate complex eigenvalues with zero real parts. The system will thus relax

on to the $2(n-1)$ dimensional center manifold spanned by the eigenvectors associated to the eigenvalues $\lambda_{c_{k,1/2}}$ and the dynamics there will be determined by the nonlinear dynamics of the system.

To gain some understanding we expand these eigenvalues for small mutation rates $\mu_x, \mu_y$ to arrive at:

$$\lambda_{s_1} = -\alpha\left(1-\frac{1}{n}\right) + \mu_x L(\alpha+1) - L\mu_y\frac{\alpha}{n}\left(\frac{n}{\gamma}+1\right) + O(\mu_x^2, \mu_y^2, \mu_x\mu_y),$$

$$\lambda_{s_2} = -\frac{\gamma}{2} + \mu_x L\gamma\frac{1+\frac{1}{\alpha}}{n-1} - L\mu_y\left(\frac{\gamma}{n}+1\right) + O(\mu_x^2, \mu_y^2, \mu_x\mu_y),$$

$$\lambda_{c_{k,1}} = \bar{\lambda}_{c_{k,2}} = i\sqrt{\alpha\gamma}\left[\frac{1}{n} - \frac{L}{n-1}\left\{\frac{\mu_x}{2}\left(1+\frac{1}{\alpha}\right) + \frac{\mu_y}{2}\left(2\left(1-\frac{1}{n}\right) + \frac{1}{\gamma}(n-1)\right)\right\}\right] + O(\mu_x^2, \mu_y^2, \mu_x\mu_y). \quad (S11)$$

We see that the oscillation speed, to leading order, is given by $\frac{\sqrt{\alpha\gamma}}{n}$, while the decay to the center manifold is governed by $\alpha$ and $\gamma$.

### 2. Calculation of eigenvectors

As shown in the last section, the system has a $2(n-1)$ dimensional center manifold. It is thus essential to incorporate the effects of nonlinear terms in order to determine stability properties of the coexistence fixed point, which will involve the diagonalization of the Jacobian. For $n=2$ it is possible to calculate analytically a linear transformation $\mathbf{T}$, which diagonalizes the Jacobian of the system of equations (S6) evaluated at the coexistence fixed point $x_p = x_q = x_m$ and $y_p = y_q = y_m$ (cf. Eqs. (S8). To make this expression analytically tractable for the later calculations, we approximated the transformation for small mutation rates as the Rayleigh-Schrödinger perturbation [8] of the Jacobian up to first order in $\mu_x$ and $\mu_y$. The four approximated eigenvalues of the Jacobian at the fixed point are computed as

$$\lambda_{s_1} = -\frac{\alpha}{2} + L(\alpha+1)\mu_x - L\frac{\alpha}{2}\left(\frac{2}{\gamma}+1\right)\mu_y < 0,$$

$$\lambda_{s_2} = -\frac{\gamma}{2} + L\gamma\left(\frac{1}{\alpha}+1\right)\mu_x + L\left(1+\frac{\gamma}{2}\right)\mu_y < 0,$$

$$\lambda_{c_1} = \bar{\lambda}_{c_2} = i\frac{\sqrt{\alpha\gamma}}{2}\left(1 - L(\mu_x+\mu_y) - L\mu_x\frac{1}{\alpha} - L\mu_y\frac{1}{\gamma}\right), \quad (S12)$$

where terms $O(\mu_x^2, \mu_y^2, \mu_x\mu_y)$ have been omitted and the bar indicates the complex conjugate. Note that this result calculated from perturbation theory coincides exactly with the series expansion of Eqs. (S11) for $n=2$, thereby validating this approach. The advantage of using the perturbation theory lies in the calculation of approximated eigenvectors, which give the transformation $\mathbf{T}$ we will need in the following.

For the sake of simplicity, the main text frequently uses $\alpha = \gamma$. Since in this case the unperturbed Jacobian (for $\mu_x = \mu_y = 0$) has two degenerate eigenvalues equal to $\frac{\alpha}{2}$, the perturbation theory for this choice of parameters results in a different transformation [8] and thus slightly different approximated eigenvalues. For completeness we give these eigenvalues here, and note that in the following they will lead to exactly the same results as the more general theory for $\alpha \neq \gamma$.

$$\lambda_{s_{1/2}}^{\alpha=\gamma} = -\frac{\alpha}{2} + L(\alpha+1)\mu_x \pm \frac{L}{2}\sqrt{(\alpha+2)^2\mu_y^2 - 4(\alpha+1)\mu_x\mu_y} < 0,$$

$$\lambda_{c_1}^{\alpha=\gamma} = \bar{\lambda}_{c_2}^{\alpha=\gamma} = i\left(\frac{\alpha}{2} - \frac{L}{2}(\mu_x+\mu_y)(\alpha+1)\right). \quad (S13)$$

### 3. Transformation to normal form

As stated in the main text, we restrict the following analysis to the case of $n=2$. Introducing the notation $\mathbf{p} = (x_p, x_q, y_p, y_q)^T$ and $\mathbf{p_m} = (x_m, x_m, y_m, y_m)^T$, we denote the corresponding coordinates of the eigensystem as functions of the original coordinates (shifted to the fixed point) by

$$(s_1, s_2, c_1, c_2)^T = \mathbf{T^{-1}}(\mathbf{p} - \mathbf{p_m}).$$

Here, $\mathbf{s} = (s_1, s_2)^T$ are the coordinates corresponding to the two negative eigenvalues $\lambda_{s_{1/2}}$ (the stable manifold) and $\mathbf{c} = (c_1, c_2)^T$ are the coordinates corresponding to the two conjugate and purely imaginary eigenvalues $\lambda_{c_{1/2}}$ (the center manifold). Note that $c_1 = \bar{c}_2$, i.e. the center manifold coordinates are complex conjugates. This yields the transformed system of stochastic differential equations for the center manifold :

$$d\mathbf{c} = [\mathbf{A}\mathbf{c} + \mathbf{f}(\mathbf{c}, \mathbf{s})]\,dt + \frac{1}{\sqrt{N}}\mathbf{D}(\mathbf{c}, \mathbf{s})d\mathbf{V}. \tag{S14}$$

Here, $\mathbf{f}$ is a nonlinear function, $\mathbf{A}$ is the diagonal matrix with entries $(\lambda_{c_1}, \lambda_{c_2})$ and $\mathbf{V}$ a two dimensional Wiener process with zero mean and unit variance. $\mathbf{D}$ is defined by the new noise covariance matrix $\mathbf{B} = \mathbf{D}^T\mathbf{D}$, which can be calculated by the Itô chain rule [4] from the old diagonal covariance matrix $\mathbf{C}$ with entries $(C_{11}^x, C_{22}^x, C_{11}^y, C_{22}^y)$ [9] as:

$$D_{ij} = (\nabla c_i(x_p, x_q, y_q, y_q))^T \mathbf{C} (\nabla c_j(x_p, y_p, y_p, y_q)).$$

As above, the derived expressions for all coefficients are valid only for small mutation rates, so terms $O(\mu_x^2, \mu_y^2, \mu_x\mu_y)$ are dropped. Finally, the dependence on the variables $\mathbf{s}$ can be removed by applying the center-manifold theorem [10]. This yields a parametrization $\mathbf{s} = \mathbf{h}(\mathbf{c})$, which reduces Eqs. (S14) to a closed two-dimensional system of conjugate complex stochastic differential equations. Keeping the notation the same, we give here only the functions and coefficients for the case $\alpha = \gamma$, since the general case is rather lengthy and does not yield any particular insight. These are (note that $c_2 = \bar{c}_1$):

$$d\mathbf{c} = [\mathbf{A}\mathbf{c} + \mathbf{f}(\mathbf{c})]\,dt + \frac{1}{\sqrt{N}}\mathbf{C}(\mathbf{c})d\mathbf{V}, \tag{S15}$$

$$f_1(c_1, c_2) = \frac{2}{5}\{L\left[c_1^3(3+i) + c_1c_2^2(1+3i) + c_1^2c_2(1+2i)(1+i) + c_2^3(1-2i)(1+i)\right]\cdot[(\alpha+1)\mu_x - \mu_y]$$
$$+ \alpha\left[c_1^2c_2(1+2i)(2+i) + c_2^3(1-2i)(2-i)\right]\},$$

$$f_2(c_1, c_2) = \overline{f_1(c_1, c_2)},$$

$$B_{11} = B_{22} = c_1^4 L\left(\frac{1}{5} + \frac{2}{5}i\right)[(\alpha+1)\mu_x - \mu_y] + c_2^4 L\left(\frac{1}{5} - \frac{2}{5}i\right)[(\alpha+1)\mu_x - \mu_y]$$
$$+ c_1^2\left(-i\frac{\alpha}{4} + L\left(\frac{1}{10} - \frac{1}{20}i\right)\left[\mu_x\left((1+4i)(\alpha+1) + \frac{4}{\alpha} + 4\right) + \mu_y\left(\frac{4}{\alpha} + 3\right)\right]\right)$$
$$+ c_2^2\left(i\frac{\alpha}{4} + L\left(\frac{1}{10} + \frac{1}{20}i\right)\left[\mu_x\left((1-4i)(\alpha+1) + \frac{4}{\alpha} + 4\right) + \mu_y\left(\frac{4}{\alpha} + 3\right)\right]\right)$$
$$- \frac{2}{5}c_1^2c_2^2 L\left[(\alpha+1)\mu_x + \mu_y\right] + \frac{1}{16}\left[L(\alpha+1)(\mu_x + \mu_y) - \alpha\right]$$

$$B_{12} = B_{21} = -c_1^4 L\left(\frac{1}{5} + \frac{2}{5}i\right)[(\alpha+1)\mu_x + \mu_y] - c_2^4 L\left(\frac{1}{5} - \frac{2}{5}i\right)[(\alpha+1)\mu_x + \mu_y]$$
$$+ c_1^2\left(-i\frac{\alpha}{4} - L\left(\frac{1}{10} - \frac{1}{20}i\right)\left[\mu_x\left(7\alpha + \frac{4}{\alpha} + 11\right) - \mu_y\left(\frac{4}{\alpha} + 3\right)\right]\right)$$
$$+ c_2^2\left(i\frac{\alpha}{4} - L\left(\frac{1}{10} + \frac{1}{20}i\right)\left[\mu_x\left(7\alpha + \frac{4}{\alpha} + 11\right) - \mu_y\left(\frac{4}{\alpha} + 3\right)\right]\right)$$
$$+ c_1^2c_2^2\frac{2}{5}L\left[(\alpha+1)\mu_x + \mu_y\right] + \frac{1}{16}\left(4 + 3\alpha - L\left[\mu_x\left(7\alpha + \frac{4}{\alpha} + 11\right) + \mu_y\left(3\alpha + \frac{4}{\alpha} + 7\right)\right]\right)$$

We continue by transforming the system into polar coordinates $u = c_1 \cdot c_2$ (squared radius) and $\varphi = \frac{1}{2i}\log\frac{c_1}{c_2}$. From the latter definition it is quite straightforward to derive a differential equation of the phase variable $\varphi$ in the deterministic limit:

$$\frac{d\varphi}{dt} = \frac{\sqrt{\alpha\gamma}}{2} - \frac{L}{2}\sqrt{\frac{\gamma}{\alpha}}(\alpha+1)\mu_x - \frac{L}{2}\sqrt{\frac{\alpha}{\gamma}}(\gamma+1)\mu_y - u\cdot\sqrt{\frac{\alpha}{\gamma}}\left((\alpha+\gamma) + 2L\frac{(\alpha+1)(\alpha+2\gamma)}{\alpha+4\gamma}\mu_x + 2L\frac{(2\alpha+\gamma)}{4\alpha+\gamma}\mu_y\right) + g(\varphi). \tag{S16}$$

where $g(\varphi)$ is a function containing terms proportional to $\exp(i\varphi)^k, k \in \pm\{2, 4\}$, which will drop out after stochastic averaging (see below).

This term reduces for $\alpha = \gamma$ to the same result as the one calculated from the slightly differing approximated eigensystem for this case (see section II C 2):

$$\frac{d\varphi}{dt} = \frac{\alpha}{2} - \frac{L}{2}(\alpha+1)(\mu_x + \mu_y) - u \cdot 2\left(\alpha + \frac{3}{5}L\left[\mu_x(\alpha+1) + \mu_y\right]\right) + g(\varphi), \tag{S17}$$

Moving on to the radial variable for $\alpha = \gamma$, according to the Itô chain rule [4] Eq. (S15) transforms to:

$$du = [a(u,\varphi) + b(u,\varphi)]\,dt + \sqrt{D_u(u,\varphi)}dW, \tag{S18}$$

$$a(u,\varphi) = u^2\left(-a_1 + g(\varphi)\right),$$

$$b(u,\varphi) = \frac{a_2}{N} + u^2 L \frac{2}{5N}\left[\mu_x(\alpha+1) + \mu_y\right] + g(\varphi),$$

$$D(u,\varphi) = u\frac{2a_2}{N} + u^2 L \frac{1}{5N}\left[\mu_x\left(\frac{4}{\alpha} + 7 + 3\alpha\right) + \mu_y\left(\frac{4}{\alpha} + 3\right)\right] + g(\varphi),$$

$$a_1 = \frac{4}{5}L\left[(\alpha+1)\mu_x + \mu_y\right], \tag{S19}$$

$$a_2 = \frac{1}{16}\left[4 + 3\alpha - \mu_x L\left(\frac{4}{\alpha} + 11 + 7\alpha\right) - \mu_y L\left(\frac{4}{\alpha} + 7 + 3\alpha\right)\right], \tag{S20}$$

where $g(\varphi)$ again are terms proportional to $\exp(i\varphi)^k, k \in \pm\{2,4\}$ and $W$ is a Wiener process with zero mean and unit variance.

For $\alpha \neq \gamma$ we arrive at the following, more general expressions for $a_1, a_2$:

$$a_1 = 4\alpha L\left(\frac{\alpha+1}{\alpha+4\gamma}\mu_x + \frac{\mu_y}{4\alpha+\gamma}\right), \tag{S21}$$

$$a_2 = \frac{1}{16}\left[\left(\frac{2}{\alpha}+3\right)\gamma + 2 - \mu_x L\frac{(\alpha+1)(\alpha(7\gamma+2)+2\gamma)}{\alpha^2} - \mu_y L\frac{(\gamma+1)(\alpha(3\gamma+2)+2\gamma)}{\alpha\gamma}\right]. \tag{S22}$$

The coefficients of the higher order terms $u^2$ also change, but will not be stated since they are unused in the following. Simple substitution shows that these expressions reduce to the corresponding simpler expressions given in Eqs. (S19) and (S20) for the special case $\alpha = \gamma$.

## III. STOCHASTIC ANALYSIS

### A. Separation of timescales & stochastic averaging

As discussed in the main text, the (stochastic) differential equations (S17) and (S18) admit a separation of time scales. As is evident from the deterministic terms, $\varphi$ evolves on $\mathcal{O}(1)$, while $u$ shows slow geometric decay $\dot{u} \propto -u^2$. This is also evident in simulations, where changes in amplitude happen over the course of several oscillation cycles. We can therefore assume $u$ to be stationary during one oscillation cycle, and average all coefficients over one period of $\varphi$ (see [9] for an extended discussion). More precisely, we integrate the coefficients as $\frac{1}{2\pi}\int_0^{2\pi} d\varphi$.

The functions $g(\varphi)$ above are introduced as symbolic placeholders for functions consisting of linear combinations of terms proportional to $\exp(i\varphi)^k, k \in \pm\{2,4\}$. These are assumed to be the only dependencies on $\varphi$. Thus, for these functions it holds that $\frac{1}{2\pi}\int_0^{2\pi} g(\varphi)d\varphi = 0$. Further, all terms independent of $\varphi$ will be left unchanged.

After integration, Eqs. (S17) and (S18) to leading order read as follows (for $\alpha \neq \gamma$):

$$\frac{d\varphi}{dt} = \frac{\sqrt{\alpha\gamma}}{2} - \frac{L}{2}\sqrt{\frac{\gamma}{\alpha}}(\alpha+1)\mu_x - \frac{L}{2}\sqrt{\frac{\alpha}{\gamma}}(\gamma+1)\mu_y,$$

$$du = \left[-a_1 u^2 + \frac{a_2}{N}\right]dt + \sqrt{2u\frac{a_2}{N}}dW. \tag{S23}$$

The coefficients $a_1$ and $a_2$ are given in Eqs. (S19) and (S20), or Eqs. (S21) and (S22) for $\alpha \neq \gamma$. This system corresponds [4] to the Fokker-Planck equation given in Eq. (4) of the main text. Note that for $N \to \infty$ this reduces to the deterministic system Eq. (3) of the main text.

Finally, instead of the ad hoc transformation to polar coordinates, it is also possible to analytically derive nonlinear coordinate transformations that reduce the system (S15) to the normal form of a Poincaré-Andronov-Hopf bifurcation

(see e.g. [10]). This rather lengthy calculation yields the same coefficients for the deterministic system as the polar coordinates, along with additional correction terms $\dot{\varphi} \propto u^2$ for the frequency equation (S17) (verified by simulation results, not shown).

### B. Equilibrium distribution

The stochastic differential equation (S23) corresponds [4] to the Fokker-Planck equation:

$$\partial_t P(u,t) = -\partial_u [\alpha(u) P(u,t)] + \frac{1}{2} \partial_u^2 [D(u) P(u,t)] \tag{S24}$$
$$\text{with} \quad \alpha(u) = -a_1 u^2 + \frac{a_2}{N} \quad \text{and} \quad D(u) = 2\frac{a_2}{N} u$$

Assuming the radius can grow without bounds, this admits an equilibrium distribution. We set Eq. (S24) to zero, and integrate twice, which yields:

$$\log P(u) = \int \frac{2\alpha(u) - \partial_u D(u)}{D(u)} du = -\frac{Na_1}{2a_2} u^2 + c \Rightarrow P(u) = \mathcal{N} \exp\left(-\frac{Na_1}{2a_2} u^2\right),$$

where $\mathcal{N}$ is an integration constant that can be chosen to normalize the distribution $P(u)$ to one. This yields $\mathcal{N} = \sqrt{\frac{2Na_1}{a_2 \pi}}$. The finite expectation value of this distribution can be calculated as:

$$\langle u \rangle = \int_0^\infty u P(u) du = \sqrt{\frac{2}{\pi} \frac{a_2}{a_1 N}}, \tag{S25}$$

which is the value given in the main text.

### C. Mean extinction time

According to [11] the mean first passage time from $u = 0$ to $u = u_{\max}$ is given by

$$T = 2 \int_0^{u_{\max}} \frac{dy}{\psi(y)} \int_0^y dz \frac{\psi(z)}{D(z)},$$

with

$$\psi(x) = \exp\left(\int_0^x dt \frac{2\alpha(t)}{D(t)}\right).$$

We begin by calculating $\psi$:

$$\int_0^x dt \frac{2\alpha(t)}{D(t)} = \lim_{\epsilon \to 0} \int_\epsilon^x dt \left(\frac{1}{t} - N\frac{a_1}{a_2} t\right) = \lim_{\epsilon \to 0} \left(\log \frac{x}{\epsilon}\right) - N\frac{a_1}{a_2} \frac{x^2}{2}$$
$$\psi(x) = \exp\left(-N\frac{a_1}{a_2} \frac{x^2}{2}\right) \lim_{\epsilon \to 0} \frac{x}{\epsilon}.$$

We proceed by calculating the inner integral of $T$

$$\int_0^y dz \frac{\psi(z)}{D(z)} = \lim_{\epsilon \to 0} \frac{1}{\epsilon} \frac{N}{2a_2} \int_0^y dz \exp\left(-N\frac{a_1}{a_2} \frac{z^2}{2}\right) = \lim_{\epsilon \to 0} \frac{1}{\epsilon} \sqrt{\frac{N}{2a_1 a_2}} \frac{\sqrt{\pi}}{2} \text{Erf}\left(\sqrt{\frac{Na_1}{2a_2}} y\right),$$

which finally gives us the mean first passage time

$$T = \lim_{\epsilon \to 0} \sqrt{\pi} \sqrt{\frac{N}{2a_1 a_2}} \int_0^{u_{\max}} dy \frac{1}{y} \text{Erf}\left(\sqrt{\frac{Na_1}{2a_2}} y\right) \exp\left(N\frac{a_1}{a_2} \frac{y^2}{2}\right)$$
$$= \sqrt{\pi} \sqrt{\frac{N}{2a_1 a_2}} \int_0^{\sqrt{\frac{Na_1}{2a_2}} u_{\max}} dx \frac{1}{x} \text{Erf}(x) \exp(x^2)$$
$$= u_{\max} \frac{N}{a_2} {}_2F_2\left(1/2, 1; 3/2, 3/2; \frac{Na_1}{2a_2} u_{\max}^2\right).$$

Where $_2F_2(1/2, 1; 3/2, 3/2; x)$ is the generalized hypergeometric function. We define

$$\bar{N} = \frac{N}{N^*} \text{ with } N^* = \frac{2a_2}{a_1 u_{\max}^2} \quad \text{and} \quad \bar{T} = \frac{T}{T^*} \text{ with } T^* = \frac{2}{a_1 u_{\max}}$$

to arrive at the universal expression

$$\bar{T} = \bar{N} \,_2F_2\left(\frac{1}{2}, 1; \frac{3}{2}, \frac{3}{2}; \bar{N}\right).$$

The scaling of $N^*$ and $T^*$ depends on the parameters of the system as well as the maximal amplitude of oscillations $u_{\max}$ which represents the boundary at which oscillations lead to the extinction of one of the involved sequences.

The expression for $u_{\max}$ given in the main text is derived from the value of the nonlinearly transformed radial variable as a function of the concentration on the virulent virus strains and corresponding immune system sequences (to first order in the mutation rates):

$$u(x_p, x_q, y_p, y_q) = \frac{\gamma(x_p - x_q)^2 \left(2\alpha\gamma + \gamma(\alpha+1)L\mu_x - \alpha(\gamma+1)L\mu_y\right)^2 + \alpha(y_p - y_q)^2 \left(2\alpha\gamma - \gamma(\alpha+1)L\mu_x + \alpha(\gamma+1)L\mu_y\right)^2}{64\alpha^3 \gamma^2}.$$
(S26)

As can be seen from the differences, this variable becomes maximal if the distribution in each population is maximally asymmetric, e.g. if $x_p = x_{\max}, x_q = 0$ and correspondingly $y_p = y_{\max}, y_q = 0$. The maximal values under the given mutational loads (and a fitness advantages of $\alpha$, as assumed in the main text) $x_{\max}$ and $y_{\max}$ can be determined from standard quasispecies theory as $x_{\max} = \frac{Q_x(\alpha+1)-1}{\alpha}$ and $y_{\max} = \frac{Q_y(\alpha+1)-1}{\alpha}$. For small mutation rates and large $\alpha$ and $\gamma$ this yields:

$$u_{\max} \simeq \frac{\alpha + \gamma}{16\alpha} + \frac{L\left(\alpha(\alpha - 3\gamma)\mu_y - \gamma(\alpha + \gamma)\mu_x\right)}{16\alpha^2 \gamma} - \frac{L(\alpha+\gamma)(\mu_x + \mu_y)}{16\alpha}. \tag{S27}$$

However, in the given scenario of oscillations around the coexistence fixed point (see Eq. (S8)) $x_p = x_q = x_m$ one virulent virus strain, e.g. $x_q$, hits a zero concentration if the other strain has the value of $x_{\max} = 2x_m$ (assuming $x_p + x_q = $ const., which is a valid assumption under the error-tail approximation). Similarly, the maximal asymmetry during oscillations in the immune system population is, e.g., $y_p = y_{\max} = 2y_m, y_q = 0$. Using these values in Eq. (S26) gives a correction factor of $u_{\max,\text{corr}} = \beta u_{\max}$ with values $\beta$ depending on the mutation rate, ranging between $\approx 0.91 - 0.99$. These are then used to rescale the curves in the plots of Fig. 3.

For the numerical estimation of the time to transition from regular adaptive cycles to intermittent switching in simulations, we initialized the populations equally and fully localized ($x_p = x_q = y_p = y_q = 0.5$) and recorded the time at which one of the virulent virus sequences was lost from the population and the immune system stabilized at the sequence corresponding to the other virulent sequence (i.e. the equilibrium between intermittent adaptive shifts of the population composition). This process was repeated for the indicated combinations of mutation probabilities and system sizes for between 80 and 360 times (small sample sizes are due to excessively long simulation times). The data points in plots in Fig. 3 show the mean escape times, with error bars indicating the 0.95 confidence interval of the mean.

### D. Limits

Expanding the expected oscillation amplitude $\langle u \rangle = \sqrt{\frac{2}{\pi} \frac{a_2}{a_1 N}}$ for large values of $\alpha, \gamma$ as well as small mutation rates, we can approximate

$$\langle u \rangle \simeq \frac{1}{4}\sqrt{\frac{3}{2\pi} \frac{\gamma(\alpha + 4\gamma)}{\alpha^2 L \mu_x N}}. \tag{S28}$$

Similarly, with $N^* = 2\frac{a_2}{a_1 u_{\max}^2}$ and $T^* = \frac{2}{a_1 u_{\max}}$, we get

$$N^* \simeq \frac{24\gamma(\alpha + 4\gamma)}{(\alpha + \gamma)^2 L \mu_x} \tag{S29}$$

$$T^* \simeq \frac{8(\alpha + 4\gamma)}{\alpha(\alpha + \gamma)\mu_x L}. \tag{S30}$$

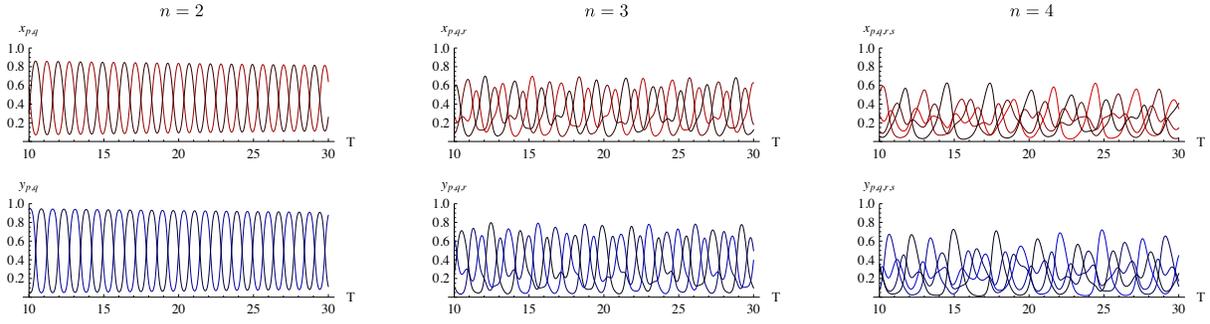

FIG. S1: Deterministic solutions of Eqs. (2) of the main text for $n = 2$ (left), $n = 3$ (middle), or $n = 4$ virulent strains (right), respectively. Simple oscillations for $n = 2$ are replaced by more complex yet periodic patterns for $n > 2$. Parameters are $\alpha = 10$, $\gamma = 15$, $\mu_x = 0.005$, $\mu_y = 0.001$ and $L = 8$.

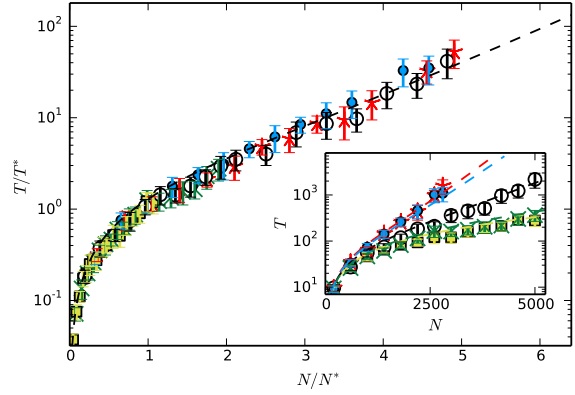

FIG. S2: Mean time until transition from regular oscillations to intermittency for varying mutation rates and increased $\gamma = 1.5\alpha$. Rescaling correctly reproduces the effects of increased $\gamma$, since the curves still collapse to the universal curve. Inset shows unscaled waiting times which are reduced w.r.t. Fig. 3 of the main text.

## IV. SUPPLEMENTAL FIGURES


\end{document}